\documentclass[useAMS,usenatbib]{mn2e} \usepackage{graphicx}

\title[QM approach to Zeeman magnetic field orientation]{A quantum mechanical approach to establishing the magnetic field orientation from a maser Zeeman profile}

\author[Green et al.]
       {J. A. Green\thanks{E-mail:j.green@skatelescope.org}$^{1,2}$, M. D. Gray$^{3}$, T. Robishaw$^{4}$, J. L. Caswell$^{2}$,\newauthor  and N. M. McClure-Griffiths$^{2}$\\
$^{1}$SKA Organisation, Jodrell Bank Observatory, Lower Withington, Macclesfield, SK11 9DL, UK\\
$^{2}$CSIRO Astronomy and Space Science, Australia Telescope National Facility, PO Box 76, Epping, NSW 1710, Australia\\
$^{3}$Jodrell Bank Centre for Astrophysics, Alan Turing Building, School of Physics and Astronomy, University of Manchester,\\ Manchester, M13 9PL, UK \\ 
$^{4}$National Research Council Canada, Herzberg Astronomy and Astrophysics Programs, Dominion Radio Astrophysical Observatory,\\ PO Box 248, Penticton, BC V2A 6J9, Canada }
\date{Accepted 2014 February 28. Received 2014 February 13; in original form 2013 August 7}

\pagerange{\pageref{firstpage}--\pageref{lastpage}} \pubyear{XXXX}
\begin{document} \maketitle

\label{firstpage}

\begin{abstract}
{Recent comparisons of magnetic field directions derived from maser Zeeman splitting with those derived from continuum source rotation measures have prompted new analysis of the propagation of the Zeeman split components, and the inferred field orientation.  In order to do this, we first review differing electric field polarization conventions used in past studies. With these clearly and consistently defined, we then show that for a given Zeeman splitting spectrum, the magnetic field direction is fully determined and predictable on theoretical grounds: when a magnetic field is oriented away from the observer, the left-hand circular polarization is observed at higher frequency and the right-hand polarization at lower frequency. This is consistent with classical Lorentzian derivations. The consequent interpretation of recent measurements then raises the possibility of a reversal between the large-scale field (traced by rotation measures) and the small-scale field (traced by maser Zeeman splitting).}
\end{abstract}


\section{Introduction}\label{introduction}
Although a large number of magnetic field studies have been undertaken using Zeeman splitting of maser spectra \citep[e.g.][]{fish05,surcis11}, the majority of these studies only consider magnetic fields for individual regions.  For mapping the field pattern within a source, the intensity of the field is of prime interest, together with changes in field direction, but knowledge of the actual line-of-sight field orientation (either towards or away from the observer) is not usually of importance to the interpretation.

However, when considering ensembles of sources, there is a possibility of comparing absolute field directions with Galactic structure, and with measurements obtained by other techniques. Results of the MAGMO survey \citep{green12magmo0}, and prior observations of magnetic field orientation from hydroxyl (OH) maser Zeeman splitting \citep[e.g.][]{reid90,fish03b,han07}, have led us to re-evaluate the field direction for a given Zeeman pattern. Specifically, we address the apparent contradiction in field direction between the maser measurements and those inferred from Faraday rotation \citep[e.g.][]{brown07,vaneck11} by exploring the Zeeman splitting in the quantum mechanical sense.

\begin{figure}
\begin{center}
\includegraphics[width=75mm]{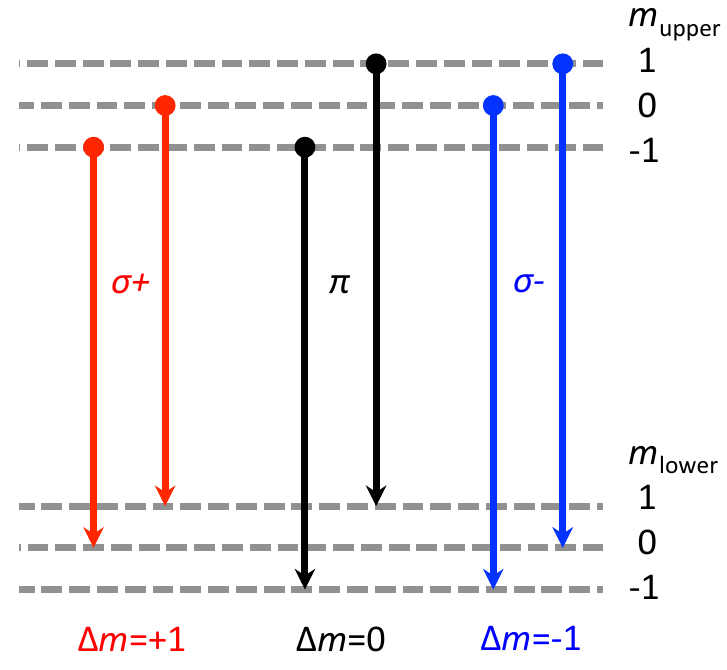}
\caption{Transitions between magnetic sub-levels of Zeeman splitting. $\Delta m = m_{\rmn{lower}}-m_{\rmn{upper}}$ \citep[e.g.][]{garcia88,gray94,gray12}.  $\Delta m = +1$ has the lower frequency (higher equivalent Doppler radial velocity), $\Delta m = -1$ has the higher frequency (lower velocity).}
\label{splittingfig}
\end{center}
\end{figure}

In the weak field limit, Zeeman splitting causes the otherwise degenerate energy levels of an atom or molecule to split into $2 {J}+1$ magnetic components, where $ {J}$ is the total angular momentum quantum number. In the simplest case of a ${J} =1-0$ transition, this results in three transition components\footnote{We focus on this simple instance, applicable to the OH doublet transitions at 1665 and 1667 MHz, and H{\sc i} at 1420 MHz. We note that similar analysis can be applied to the more complex Zeeman patterns of some other transitions, such as the 1720 MHz satellite transition of OH.}: the unshifted (in frequency relative to zero magnetic field) $\pi$ and the two shifted $\sigma$s, denoted $\sigma^{+}$  and $\sigma^{-}$ (Figure\,\ref{splittingfig}). Commonly, conventions are invoked when attributing the $\sigma^{+}$  and $\sigma^{-}$ components to a handedness of circular polarization, and for allocating which of these is found at the higher frequency for a given field direction.

In this paper we first outline the current convention for inferring field orientation from an observed maser spectrum (Section\,\ref{conventions}). We then re-evaluate the propagation of the individual components to show how the field direction is fully determined and predictable on theoretical grounds, and is consistent with the previously used convention (for example as adopted in \citealt{davies74} and \citealt{garcia88}). The argument is presented first in an abbreviated descriptive form (Section\,\ref{descriptionbrief}) before a full derivation (Section\,\ref{descriptionfull}). Furthermore, in the appendix we test the compliance of various maser theory publications that discuss polarization (considering the direction of waves, the standard Cartesian axis system and the polarization conventions).

\section{Commonly adopted magnetic field direction convention}\label{conventions}
Conventions of field orientations have a long and chequered past, exacerbated by differences between optical and radio wavelengths, emission and absorption, the pulsar community and the rest of the astronomy community \citep[e.g.][]{babcock53a,babcock53b}. The use of polarization conventions in theoretical papers over the years has similarly been inconsistent. The handedness of polarization in theoretical work is determined by the pair of helical vectors (in the spherical coordinate basis) that are used to represent
left-hand circular polarization (LCP) and right-hand circular polarization (RCP). 
A detailed history of the early measurements and conflicting conventions is given by \citet{robishaw08}.

There are essentially three elements that have to be taken into account to define the field direction: 1. what is defined as RCP and LCP polarization (invoking coordinate systems and basis vectors); 2. which $\sigma$ components these polarizations interact with; and 3. which frequencies these polarizations are found at for a field towards us or away from us. 

\begin{figure}
\begin{center}
\includegraphics[width=84mm]{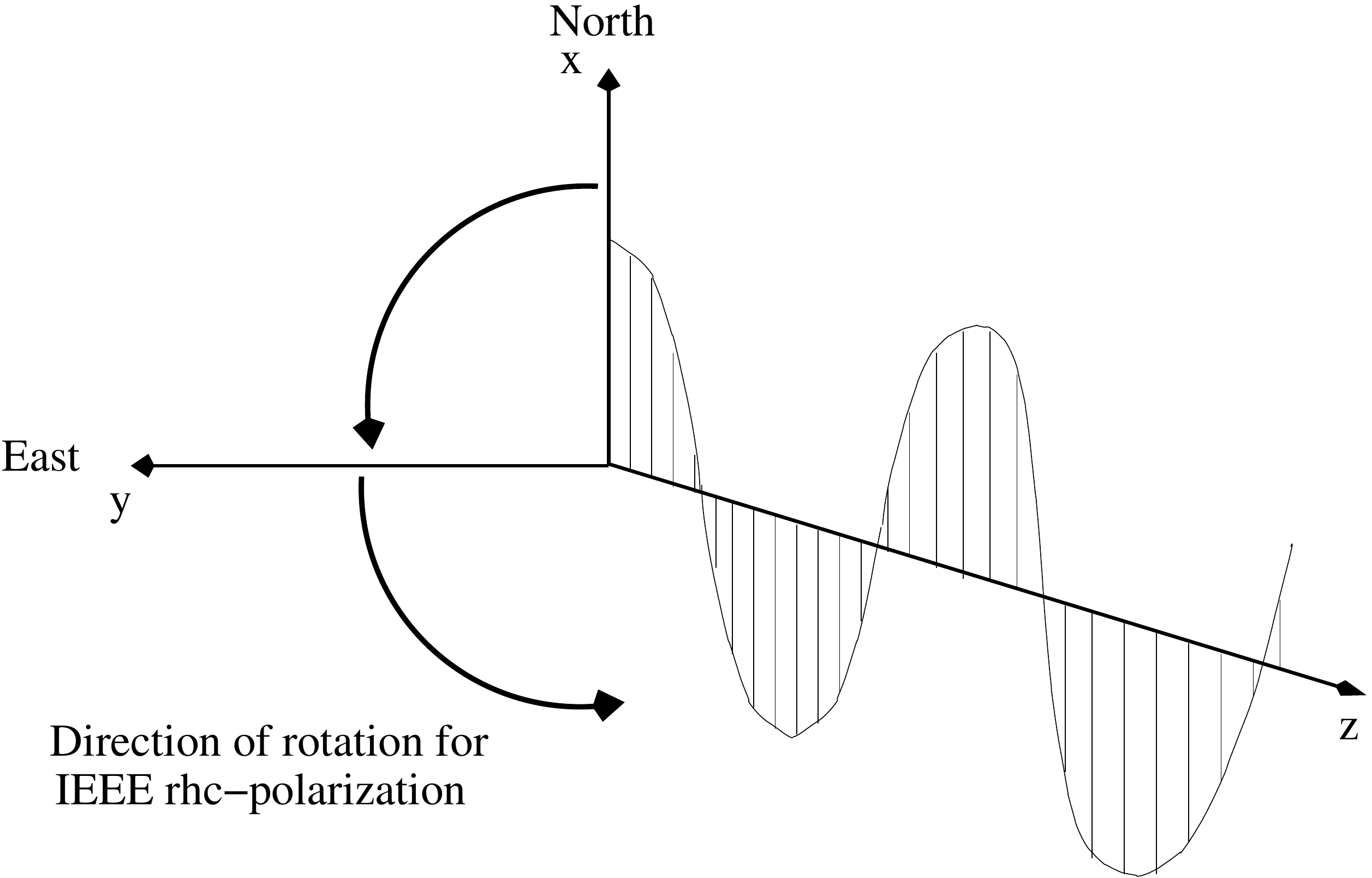}
 \caption{The `right-handed' axis system with the electric field of the wave from equation (\ref{eq:eposz}) sketched in, and showing the direction of rotation of the electric field vector, as seen by the observer, of RCP radiation under the IEEE convention. The alignment of the $x$ and $y$ axes with, respectively, North and astronomical East follows the standard IAU orientations as set out in \citep{hamaker96}.}
  \label{xyz}
  \end{center}
\end{figure}

The IEEE convention\footnote{The radio engineering definition of RCP and LCP dates from 1942 (as decreed by the IRE, Institute of Radio Engineers) but is commonly
referred to as the IEEE standard (endorsed in 1969 by the IEEE, which had been formed in 1963 as a merger of IRE and IEE).} is the current standard for the first element, defining LCP as clockwise rotation of the electric field vector as viewed by the observer 
with radiation approaching, and RCP as counterclockwise (see also Figure~\ref{xyz}). Radio astronomers adopted the IEEE usage, and it was formally endorsed in 1973 by the IAU (Commission 40 chaired by G. Westerhout). Unfortunately, an opposite widely used convention is adopted in classical optics, by both physicists and optical astronomers. Tested sets of helical vectors in later sections may therefore be described as either IEEE-compliant or optics-compliant.

 For the next two elements, we consider both observations and the IEEE convention for Stokes $V$. The definition of Stokes $V$ is required for field directionality as discussed later. The IAU convention is that: Stokes $V$ is RCP minus LCP, therefore RCP corresponds to positive $V$ and LCP to negative $V$, i.e.
\begin{equation}
V = ({\rm RCP} - {\rm LCP}) = (\tilde{\cal E}_R \tilde{\cal E}_R^*
                -\tilde{\cal E}_L \tilde{\cal E}_L^*),
\label{eq:IAUStokesV}
\end{equation}
\noindent
the second expression being the representation in terms of electric field amplitudes of the two polarizations as helical vectors in the spherical basis  \citep[e.g.][]{landau82}. The tilde indicates a complex-valued function and the asterisk the complex conjugate.

In order to apply these conventions to observations, it is also necessary to know whether an observed shift of LCP to lower frequency, i.e. equivalent higher Doppler radial velocity (and RCP to higher frequency, or lower velocity) corresponds to a field oriented towards or away from the observer.  An early paper where this is an issue of special interest is \citet{davies74}, where the field direction for a group of sources is compared to the direction of Galactic rotation.  That paper asserts that RCP shifted to higher velocity (as in the case of the much studied W3(OH) region) corresponds to a field away from the observer. The paper also describes this field orientation as a positive magnetic field.  These are the same conventions used in earlier papers considering H{\sc i} absorption \citep{davies62,verschuur69}.   All subsequent papers that we are aware of, and in particular the commonly  cited paper by  \citet{garcia88}\footnote{We note that in this paper, confusingly, the labelled Stokes $V$ has a sign inconsistent with the IAU definition of $V$.} have also retained this convention.  None of the papers show a derivation justifying this convention, and the later papers in particular have merely adopted the convention without reassessing if it is correct.

However, accepting the above assertion, or convention, the magnetic field orientation from Zeeman splitting of maser emission for the Carina-Sagittarius spiral arm is found to be opposite to that indicated by rotation measures of Galactic and extragalactic sources \citep[][and references therein]{green12magmo0}. It is this apparent discrepancy that prompted a rigorous re-evaluation. 

\subsection{The classical Lorentzian derivation}
Although there has been much ambiguity within the astronomical community, there is a significant body of physics literature with derivations, in a classical sense, of the field direction from Zeeman splitting. These started with the original work by \citet{zeeman1897,zeeman1913} and include: \citet{white34,sommerfeld54,stone63,jenkins76,landi04,haken05}. In this work if the magnetic field is directed towards the observer (and denoted with a negative value), IEEE LCP is at the lower frequency, IEEE RCP is at the higher frequency. Similarly if the magnetic field is directed away from the observer (and denoted with a positive value) IEEE LCP is at the higher frequency, IEEE RCP is at lower frequency. Throughout the rest of the paper we refer to this body of work as the `classical Lorentzian derivation'.

\section{The inferred field direction}\label{descriptionbrief}
In this section, we revisit the quantum mechanics and radiative transfer of the Zeeman effect to demonstrate that the inferred direction of the field is uniquely defined by the observed frequency (or velocity) shift, in accordance with the classical Lorentzian derivation.

\subsection{Frequencies of $\sigma$ components}\label{sigfreq}
\begin{figure}
\begin{center}
\renewcommand{\baselinestretch}{1.1}
\includegraphics[width=8cm]{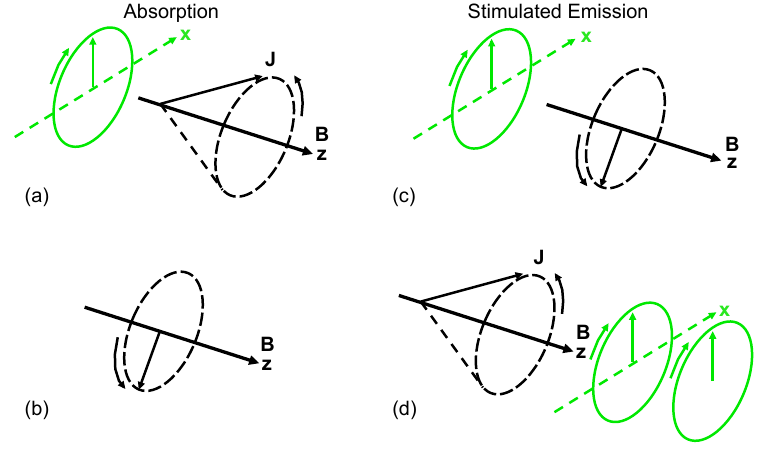}  
\caption{\small Radiation (green dashed lines and solid circles) and molecule (black solid lines and dashed circles) interaction in a $\sigma^+$ transition. The left side represents the interaction for absorption of radiation, with an IEEE right-hand rotating molecule absorbing LCP radiation, the right side represents the interaction for stimulated emission, with incident LCP radiation on an unpolarized molecule resulting in a IEEE right-hand rotating molecule and twice the LCP radiation. (a) and (c) show the initial states of the interaction, (b) and (d) the final states. Radiation is propagating in the $+z$ direction.}
\label{molec}
\end{center}
\end{figure}

\begin{figure}
\begin{center}
\renewcommand{\baselinestretch}{1.1}
\includegraphics[width=8cm]{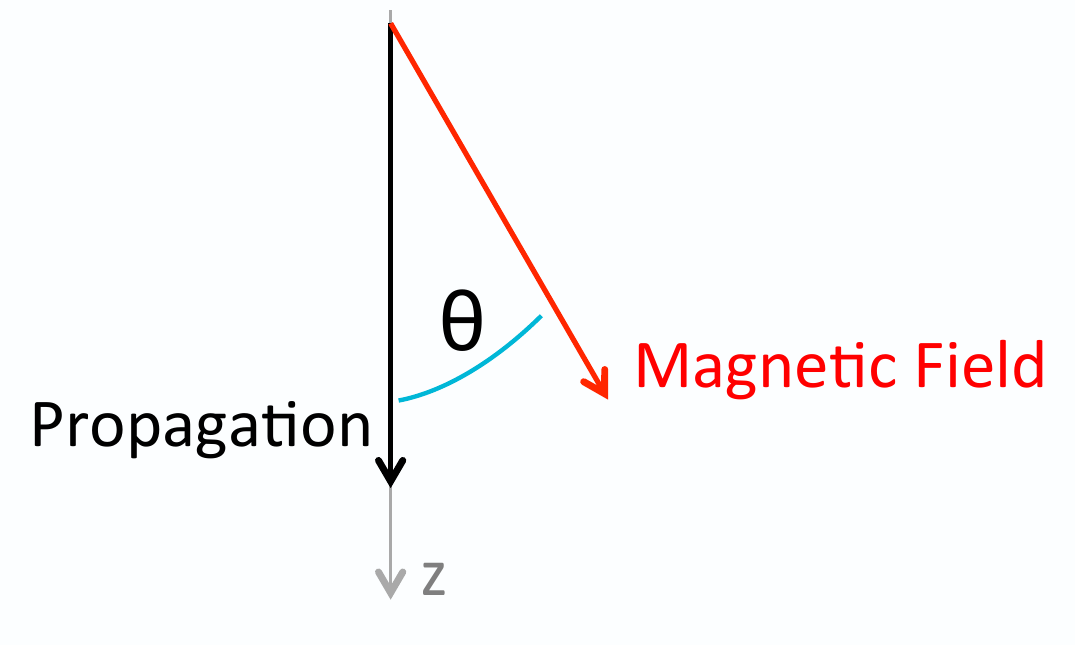}  
\caption{\small Definition of propagation and field direction vectors, $\theta$ represents the angle between the two, and the observer is looking towards the $z$-axis from below.}
\label{vectors}
\end{center}
\end{figure}

We now consider the interaction of circularly polarized radiation with molecules that are subject to Zeeman splitting by an external magnetic field (Figure \ref{molec}). According to the discussion in \citet{eisberg74}, the magnetic moment of the molecule, $\bmath{\mu}$, is close to, but not exactly, anti-parallel to the total angular momentum vector, $\bmath{J}$, (or $\bmath{F}$ in a molecule like OH or CH that has a Zeeman effect of hyperfine structure). The magnetic moment precesses rapidly about $-\bmath{J}$ and much more slowly about $-\bmath{B}$. \citet{eisberg74} introduce the approximation that in one period of rotation of $\bmath{\mu}$ about $-\bmath{B}$, $\bmath{\mu}$ will rotate so many times about $-\bmath{J}$ that the component of $\bmath{\mu}$ perpendicular to $-\bmath{J}$ averages out to zero, and we need to consider only the parallel component, $\mu_{\bmath{J}}$, precessing with $-\bmath{J}$ about $-\bmath{B}$. This precession implies a corresponding precession of
$\bmath{J}$ about $\bmath{B}$, and
it may be shown \citep[for example][]{littlefield79} that the sense of this
latter precession is counterclockwise for observers with the
magnetic field pointing towards them. Aligning the magnetic
field and radiation propagation directions along the $z$-axis
($\theta = 0$, see Figure \ref{molec}), observers receiving
the radiation also see $\bmath{J}$ rotating counterclockwise,
corresponding to right-handed rotation under the IEEE convention
(Figure~\ref{vectors}). Since $m$ is the quantum number
corresponding to $J_z$, the projection of $\bmath{J}$ on the
$z$-axis, this right-handed (counterclockwise) rotation corresponds
to positive values of $m$. 

The considerations above allow us to consider the radiation-molecule interaction in a $\sigma^+$ transition (Figure\,\ref{molec}). Recall that in our convention, a $\sigma^+$ transition is one in which the value of $m$ increases by 1 in emission. The left-hand side of the figure shows the transition in absorption. The molecules change from the initial state (a) as discussed above (IEEE right-hand rotation and $m = 1$) to the final state (b) where $\bmath{J}$ has no projection on the $z$-axis and $m =0$. The overall value of $m$ must therefore be zero. To conserve the angular momentum of the interaction, the initially right-handed molecules must interact with LCP radiation (under the IEEE convention), which has an electric field vector that rotates clockwise as viewed by an observer receiving the radiation (Figure\,\ref{molec}(a)).  A derived result of this scheme is that a photon of the LCP radiation must carry --1 unit of angular momentum associated with $m$. This result is consistent with the conventions on photon polarization in \citet{landau82} and \citet{yang10}, having taken into account the handedness conventions used in these works. On the right-hand side of Figure~\ref{molec} we see a stimulated emission event, with the LCP radiation now approaching a molecule with $m = 0$ (part (c)). A photon of this radiation then copies itself, and leaves the molecules in right-hand precession (d). The overall value of $m$ in this stimulated emission case is $-1$.  It should be noted that for the case of spontaneous emission, one can follow the left side of Figure~\ref{molec} from (b) to (a): start with $m=0$ and no radiation and the result is radiation with LCP, and the molecules have undergone a right-handed transition, increasing $m$ by 1; this is in the same sense as for stimulated emission. 

For Zeeman splitting of maser emission, if  $\Delta m = m_{\rmn{lower}}-m_{\rmn{upper}}$ \citep[e.g.][]{gray94,gray12}, where $m_{\rmn{lower}}$ and  $m_{\rmn{upper}}$ are the quantum numbers corresponding to the magnetic sub-levels (Figure \ref{splittingfig}), we find that $\sigma^{+}$ is always found at the lower frequency (higher velocity), $\sigma^{-}$ at the higher frequency (lower velocity).\footnote{This convention is presented in the often cited \citet{garcia88}, although it should be noted that the alternative  $\Delta m = m_{\rmn{lower}}-m_{\rmn{upper}}$ is also often adopted (as noted by D.~E.~Rees in \citealt{kalkofen88}).} This is shown schematically in Figure \ref{splittingfig}.

\subsection{$\sigma$ to Stokes $V$ correspondence}\label{sigcorrespond}
The evolution of Stokes $V$ with propagation distance for the two $\sigma$ components, following \citet{goldreich73part2}, hereafter GKK73, can be defined respectively for the $\sigma^{+}$  and $\sigma^{-}$ components as\footnote{ Note that the $\pm 1$ subscripts used on the Stokes parameters to denote transition type by GKK73 are reversed with respect to the $\sigma^\pm$ notation used in the present work.}:

\begin{equation}
{dV}/{dz} = \gamma(1 + cos^{2} \theta)V - 2\gamma cos \theta I
\label{eq:sigmaplus}
\end{equation}

\begin{equation}
{dV}/{dz} = \gamma(1 + cos^{2} \theta)V +2\gamma cos \theta I,
\label{eq:sigmaminus}
\end{equation}

\noindent
where $\gamma$ is the gain coefficient, and with the field vectors and $\theta$ defined as in Figures \ref{molec} and \ref{vectors}. Remembering that Stokes $I$ exceeds Stokes $V$, it can be seen from this equation that for an aligned field, one where the magnetic field vector is approximately coincident with the propagation vector ($\theta \approx 0$), thus directed towards the observer, the $\sigma^{+}$ component will have increasingly negative Stokes $V$ and the $\sigma^{-}$ component increasingly positive Stokes $V$. Similarly for an opposing field direction ($\theta \approx \pi$), the $\sigma^{+}$ component will have increasingly positive Stokes $V$ and the $\sigma^{-}$ component increasingly negative Stokes $V$. Thus, with the frequencies (or equivalent velocities) of the $\sigma^{+}$  and $\sigma^{-}$ components defined by the quantum mechanics of the splitting (Section \ref{sigfreq}), we know inherently that positive Stokes $V$ at a lower frequency (higher velocity) indicates a field directed away from the observer.

\subsection{Polarization handedness}\label{polhand}
If we now, as is commonly done, invoke the IAU definition of Stokes $V$  for emission, with positive $V$ corresponding to RCP and negative $V$ corresponding to LCP, we see that RCP at a lower frequency (higher velocity) and LCP at a higher frequency (lower velocity) indicates a field away from the observer.

\section{Consistency of polarization definitions}\label{descriptionfull}
In this section we justify sections \ref{sigcorrespond} and \ref{polhand} by demonstrating the consistency of the statement based on helical basis vectors.

\subsection{The direction of waves} 
\label{dirwav}

We consider our electromagnetic (EM) waves to propagate in the
positive $z$ direction. The electric and magnetic fields are then
confined to the $xy$-plane (Figure \ref{xyz}). The electric field of such a plane-polarized wave has a standard representation \citep{young04,lothian80,jenkins57} of:
\begin{equation}
\bmath{E}(z,t) = \hat{\bmath{x}} {\cal E}_x \cos (\omega t - k z),
\label{eq:eposz}
\end{equation}
where $\hat{\bmath{x}}$ is the unit vector along the $x$-axis, 
${\cal E}_x$ is the field amplitude, $\omega$ is its angular frequency
and $k=\omega /c$, its wavenumber. 

\subsection{The field as applied to masers}
\label{maswav}

In equation (\ref{eq:eposz}), the field amplitude is assumed to be real and constant. In maser astrophysics, we typically deal with a spectral
line composed of Fourier components distributed about a line centre frequency, $\omega_0$. The width of the line is narrow in the sense
that some width parameter, $\Delta \omega$, such as the full width at half maximum, satisfies $\Delta \omega \ll \omega_0$. We can now generalise
the field in equation (\ref{eq:eposz}) to the typical maser case by letting the rapidly oscillating trigonometric term depend on $\omega_0$, and by
introducing a slowly-varying phase factor into the amplitude of each Fourier component: see for example, \citet{menegozzi78,goldreich73part2,deguchi90,gray12}. 
The complex amplitude of the full field, $\tilde{\cal E}_x(z,t)$ then becomes an integral over frequency offset from $\omega_0$, and varies on a timescale vastly
longer than $1/\omega_0$. The field is now, 
\begin{equation}
\bmath{E}(z,t) = \Re \left\{
  \hat{\bmath{x}} \tilde{\cal E}_x(z,t) e^{ -i\omega_0(t -  z/c) }
                     \right\},
\label{eq:elinpol}
\end{equation}
where the tilde on the amplitude indicates a complex-valued function. As discussed above, the wave in equation (\ref{eq:elinpol}) is moving in the direction of more positive $z$.

For the present purpose of testing the conformity of EM radiation definitions with polarization conventions, the full complexity of equation (\ref{eq:elinpol}) is not required. Our investigations require the use of time differences of the order of $1/\omega_0$ and distances vastly shorter than any amplification or gain length. We therefore ignore the slow time and space dependence of the complex amplitude, leaving it in the form of a constant real amplitude multiplied by a constant phase factor, $e^{i \phi_x}$, that is:
\begin{equation}
\bmath{E}(z,t) = \Re \left\{
  \hat{\bmath{x}} {\cal E}_x e^{i \phi_x} e^{ -i\omega_0 (t -  z/c) }
                     \right\}.
\label{eq:elinpolsimp}
\end{equation}

\subsection{Elliptical polarization}

The wave represented by equation (\ref{eq:elinpolsimp}) is linearly polarized in the $xz$-plane. In this work we need to consider circularly and, more generally, elliptically polarized radiation. This will contain both $x$ and $y$ components of the electric field, each with its own phase factor. In general, we have:
\begin{equation}
\bmath{E}(z,t) = \Re \left\{ 
                       \left[
  \hat{\bmath{x}} {\cal E}_x e^{i \phi_x} +
  \hat{\bmath{y}} {\cal E}_y e^{i \phi_y} 
                       \right] e^{ -i\omega_0 (t -  z/c) }
                     \right\},
\label{eq:ellipcart}
\end{equation}
where $\phi_x$ and $\phi_y$ are the phases of the Cartesian field components.

A Cartesian representation is unwieldy when calculating the interaction of the the EM radiation with the molecular density matrix, so it is customary to shift to a set of unit vectors based on positive and negative helicity: helical unit vectors in the spherical basis, often written as $\hat{\bmath{e}}_+$ and $\hat{\bmath{e}}_-$. The field amplitude can now be broken into helical, rather than Cartesian, components, so that the usual representation for our elliptically polarized wave is:
\begin{equation}
\bmath{E}(z,t) = \Re \left\{
                       \left[
  \hat{\bmath{e}}_+ \tilde{\cal E}_+(z,t) +
  \hat{\bmath{e}}_- \tilde{\cal E}_-(z,t)
                       \right]
 e^{ -i\omega_0(t -  z/c) }
                     \right\}.
\label{eq:ellip_pol}
\end{equation}
There is a third helical unit vector, but this is simply equal to the $z$-axis unit vector, and is often written, 
$\hat{\bmath{e}}_0=\hat{\bmath{z}}$.
The positive and negative helicity unit vectors are unfortunately
ambiguous, and we discuss below how to attach them to a
standard pair of unit vectors corresponding to IEEE LCP
and RCP. We have removed one ambiguity by choosing to write $e^{-i\omega_0(t-z/c)}$ (rather than $e^{+i\omega_0(t-z/c)}$) when using complex exponential notation, as appears to be standard practice in maser polarization theory papers, including GKK73.

\subsection{The axis system}
\label{standardaxis}

We have defined our EM wave to propagate along the $z$-axis in the
positive direction in Section~\ref{dirwav}. This definition must be supplemented
by a convention for the orientation of the $x$ and $y$ axes if any test of
handedness is to work. We assume that the standard `right-handed'
system of axes from mathematics has been used by all authors unless they
have clearly stated otherwise. This axis system is drawn in many
textbooks, and it is reproduced in Figure~\ref{xyz}
\citep{arfken70,boas66}. Also shown in Figure~\ref{xyz} is the EM wave
from equation (\ref{eq:eposz}) at time $t=0$.

\subsection{A test prescription for polarization handedness}
\label{presc}

In mathematical descriptions of elliptically polarized radiation,
it is the helical unit vectors that decide the handedness of
polarization, given some standard definition of left and right.
Here, we present a formal prescription for testing any pair
of helical unit vectors against the IEEE standard:
\begin{enumerate}
  \item Associate the positive and negative helicity unit vectors
with presumed LCP and RCP radiation.
  \item Use equation (\ref{eq:ellip_pol}) with the presumed RCP and LCP
unit vectors to determine the RCP and LCP electric field components in
terms of their Cartesian counterparts.
  \item Write down a version of equation (\ref{eq:ellip_pol}) corresponding
to an RCP wave.
  \item Insert into this equation the definition of the presumed right-hand
unit vector, and resolve the electric field into its Cartesian components.
  \item Set a fixed distance, say $z=0$.
  \item Pick a time, $t_1$, such that the electric field is aligned
along the positive $y$-axis.
  \item Advance the time to $t_2$ so that  $\omega_0 t_2=\omega_0 t_1 +\pi/2$,
and check the new alignment of the electric field vector.
  \item If the field at $t_2$ is aligned with the negative $x$-axis, then
the field has rotated counterclockwise from the observer's point of view
(see Figure~\ref{xyz}) and the presumed RCP vector conforms
to the IEEE standard. If the field has instead rotated clockwise
from the observer's point of view to point along the positive $x$-axis,
then the presumed right-handed unit vector is actually left under the IEEE
convention (or right-handed in the optics convention).
\end{enumerate}

\subsection{Goldreich, Keeley \& Kwan (1973)}
\label{testGKK73}

The seminal theory paper on maser polarization is GKK73. Their
equation~12 clearly defines Stokes $V$ in accordance with the
IAU convention: right minus left. The same equation also tells
us that the positive helicity unit vector should be associated
with RCP (and negative helicity with LCP).
The definitions of the helical unit vectors are given in the
text just above equation~12 of GKK73 and, given the associations
above, we deduce that,
\begin{equation}
\hat{\bmath{e}}_R = (\hat{\bmath{x}} + i \hat{\bmath{y}})/\sqrt{2}
\;\; ; \;\;
\hat{\bmath{e}}_L = (\hat{\bmath{x}} - i \hat{\bmath{y}})/\sqrt{2},
\label{eq:GKK73rl}
\end{equation}
noting that in this basis the left-hand vector is the complex
conjugate of the right-hand vector. Do these vectors conform to
the IEEE definition? The short answer is yes, they do. To prove this,
we follow the prescription set out in Section~\ref{presc} above.

Step (i) of Section \ref{presc} has already been completed in the discussion above. For
step (ii), we write down a version of equation (\ref{eq:ellip_pol}) in which
the positive and negative helicity vectors and components are replaced
by their RCP and LCP equivalents:
\begin{equation}
\bmath{E}(z,t) = \Re \left\{
                       \left[
  \hat{\bmath{e}}_R \tilde{\cal E}_R +
  \hat{\bmath{e}}_L \tilde{\cal E}_L
                       \right]
 e^{ -iY }
                     \right\},
\label{eq:rlpol}
\end{equation}
where $Y(z,t)=\omega_0 (t-z/c)$. Inserting the definitions from 
equation (\ref{eq:GKK73rl}), and resolving into Cartesian components, we
obtain,
\begin{equation}
\bmath{E}(z,t) = (1/\sqrt{2}) \Re \left\{
                       \left[
      \hat{\bmath{x}} ( \tilde{\cal E}_R + \tilde{\cal E}_L ) 
  + i \hat{\bmath{y}} ( \tilde{\cal E}_R - \tilde{\cal E}_L )
                       \right]
 e^{ -iY }
                                \right\},
\label{eq:rlpol}
\end{equation}
from which it is evident that 
${\cal E}_x = (\tilde{\cal E}_R + \tilde{\cal E}_L)/\sqrt{2}$ and
${\cal E}_y =i(\tilde{\cal E}_R - \tilde{\cal E}_L)/\sqrt{2}$. Inverting
this pair of expressions requires that,
\begin{equation}
\tilde{\cal E}_L = ({\cal E}_x + i {\cal E}_y)/\sqrt{2} \;\; ; \;\;
\tilde{\cal E}_R = ({\cal E}_x - i {\cal E}_y)/\sqrt{2} .
\label{eq:GKK73fields}
\end{equation}
We can now complete step (iii) by
writing down a version of equation (\ref{eq:ellip_pol}) in
(presumed) RCP only:
\begin{equation}
\bmath{E}_R(z,t) = \Re \left\{
  \hat{\bmath{e}}_R \tilde{\cal E}_R
 e^{ -i Y }
                     \right\},
\label{eq:rpol}
\end{equation}
and continue to step (iv) by inserting the definitions of
$\hat{\bmath{e}}_R$ from equation (\ref{eq:GKK73rl}) and of
$\tilde{\cal E}_R$ from equation (\ref{eq:GKK73fields}). The result is,
\begin{eqnarray}
\bmath{E}_R(z,t) & = & (1/2) \Re \left\{
                     \left[
   \hat{\bmath{x}} {\cal E}_x + \hat{\bmath{y}} {\cal E}_y
+i(\hat{\bmath{y}} {\cal E}_x - \hat{\bmath{x}} {\cal E}_y)
                     \right]     \right. \nonumber \\
                 & \times &  \left.
                              (\cos Y - i \sin Y)
                             \right\}.
\label{eq:GKK73_inter}
\end{eqnarray}
Step (iv) is completed by multiplying out the brackets and taking the
real part, leaving
\begin{eqnarray}
\bmath{E}_R(z,t) &=& (1/2) \left\{
   \hat{\bmath{x}} ({\cal E}_x \cos Y - {\cal E}_y \sin Y)
                          \right. \nonumber \\
  &+&                   \left.
   \hat{\bmath{y}} ({\cal E}_y \cos Y + {\cal E}_x \sin Y)
                        \right\}.
\label{eq:GKK73_step4}
\end{eqnarray}
For step (v), we set $z=0$, so that $Y(0,t)=\omega_0 t$ in 
equation (\ref{eq:GKK73_step4}), and we also assume circular, rather than
elliptical, polarization, so that 
${\cal E}_x = {\cal E}_y = {\cal E}$ can be extracted as a common factor.
The electric field to test is now,
\begin{eqnarray}
\bmath{E}_R(0,t)& =& ({\cal E}/2) \left\{
   \hat{\bmath{x}} ( \cos \omega_0t - \sin \omega_0 t)
                                \right. \nonumber \\
  &+&                           \left.
   \hat{\bmath{y}} ( \cos \omega_0t + \sin \omega_0 t)
                        \right\}.
\label{eq:GKK73_step5}
\end{eqnarray}
Step (vi) is achieved by setting $\omega_0 t_1 = \pi/4$, so that
$\cos \omega_0 t_1 = \sin \omega_0 t_1 = 1/\sqrt{2}$. The $x$ component
of the field disappears, and the field is aligned with the positive
$y$-axis:
\begin{equation}
\bmath{E}_R(0,t_1) = ({\cal E}/\sqrt{2}) \hat{\bmath{y}} .
\label{eq:GKK73_step6}
\end{equation}
To see how the field rotates, we advance the time by one
quarter period to $t_2$, such that
$\omega_0 t_2 = \omega_0 t_1 + \pi/2 = 3\pi/4$. The trigonometric functions
now have the values, $\cos \omega_0 t_2 = -1/\sqrt{2}$ and
$\sin \omega_0 t_2 = +1/\sqrt{2}$. The modified field is,
\begin{equation}
\bmath{E}_R(0,t_2) = -({\cal E}/\sqrt{2}) \hat{\bmath{x}} ,
\label{eq:GKK73_step7}
\end{equation}
which completes step (vii). The final step is to note that, from
the observer's viewpoint, the field has rotated counterclockwise
through one quarter turn to align with the negative $x$-axis.
The conclusion is that the presumed RCP vector
from equation (\ref{eq:GKK73rl}) is indeed IEEE-compliant.

A similar analysis (though using a different starting time, $t_3$)
shows that the LCP vector is also IEEE-compliant
as left-handed. We conclude that the helical vectors of the spherical basis used
in GKK73 are IEEE-compliant and satisfy the IAU definition
of Stokes $V$.

\subsection{Gray \& Field (1995)}
\label{testgf95}

The first paper of a series, \citet{gray95}, denoted GF95 hereafter,
applied the semi-classical saturation theory developed in
\citet{field84} and \citet{field88}
to polarized masers, particularly for the case where the Zeeman
splitting is large compared to the Doppler width. This work is
somewhat more difficult to test than GKK73 because the electric
field definition is in Cartesian components. However, it is useful
because the helical vectors may be derived from field
and phase definitions, rather than stated.

We begin with equation (\ref{eq:ellipcart}), the definition of an elliptically
polarized wave in Cartesian components. It is straightforward to transfer
the entire phase factor to the $y$-component: lift the real part operator to
yield a complex version of the electric field, and multiply this by
$e^{-i\phi_x}$. The real part of this modified field is then the
electric field used in GF95 (their equation~1):
\begin{equation}
\bmath{E}(z,t) = \Re \left\{ 
                       \left[
  \hat{\bmath{x}} {\cal E}_x  +
  \hat{\bmath{y}} {\cal E}_y e^{-i \delta} 
                       \right] e^{ -i Y(z,t) }
                     \right\},
\label{eq:gf95_field}
\end{equation}
where $Y(z,t)$ is defined as before and $\delta = \phi_x - \phi_y$, as
stated in the text below equation~1 of GF95. If we assume circular
polarization to set ${\cal E}_x = {\cal E}_y = {\cal E}$, and expand
the complex exponentials, equation (\ref{eq:gf95_field}) may be developed
to the form,
\begin{equation}
\bmath{E}(z,t) = {\cal E} \Re \left\{ 
                       \left[
  \hat{\bmath{x}}  +
  \hat{\bmath{y}} (\cos \delta - i \sin \delta) 
                       \right] (\cos Y - i \sin Y)
                              \right\},
\label{eq:gf95_step1}
\end{equation}
and after multiplying out the brackets and taking the real part, to
\begin{equation}
\bmath{E}(z,t) = {\cal E} \Re \left\{ 
  \hat{\bmath{x}}  \cos Y +
  \hat{\bmath{y}}  \cos (\delta + Y) 
                              \right\}.
\label{eq:gf95_step2}
\end{equation}
As in the GKK73 test, we proceed by setting the fixed distance of $z=0$
to obtain
\begin{equation}
\bmath{E}(0,t) = {\cal E} \Re \left\{ 
  \hat{\bmath{x}}  \cos \omega_0 t +
  \hat{\bmath{y}}  \cos (\delta + \omega_0 t) 
                              \right\}.
\label{eq:gf95_step3}
\end{equation}
At this point, we introduce the IEEE convention\footnote{Rees in \citet{kalkofen88} uses the optics convention; his result for $\delta$ was reversed to obtain the IEEE form.}, which requires
that for RCP radiation, the $y$ component of the field
leads the $x$ component: that is $\delta$ is negative and, for
circular polarization, equal to $-\pi/2$. Inserting this value
into equation (\ref{eq:gf95_step3}), we recover
\begin{equation}
\bmath{E}(0,t) = {\cal E} \Re \left\{ 
  \hat{\bmath{x}}  \cos \omega_0 t +
  \hat{\bmath{y}}  \sin \omega_0 t
                              \right\}.
\label{eq:gf95_ieee_rhc}
\end{equation}
Note that equation (\ref{eq:gf95_ieee_rhc}) is consistent with
cosinusoidal $x$ and sinusoidal $y$ components in IEEE
(`source point-of-view').

As a final check, set the initial time $t_1$ such that
$\omega_0 t_1 = \pi/2$, and equation (\ref{eq:gf95_ieee_rhc}) reduces
to the $y$-aligned, $\bmath{E}(0,t_1) = {\cal E} \hat{\bmath{y}}$.
The time can now be advanced one quarter period, so that
$\omega_0 t_2 = \omega_0 t_1 + \pi/2 = \pi$, leading to
$\bmath{E}(0,t_2) = - {\cal E} \hat{\bmath{x}}$. This is a
counterclockwise rotation from the observer's point of view, so
the electric field in GF95 is consistent with the IEEE convention.

\subsubsection{Helical unit vector for RCP radiation}

The choices made in the test above now dictate the helical unit
vector (in the spherical basis) for RCP radiation in GF95. With $\delta=-\pi/2$ as required
for IEEE RCP, $e^{-i\delta}=i$. Substitution of this result
into equation (\ref{eq:gf95_field}), and setting 
${\cal E}_x = {\cal E}_y = {\cal E}$ as above for circular polarization,
we find that the RCP wave is
\begin{equation}
\bmath{E}_R(z,t) = {\cal E} \Re \left\{ 
                       \left[
  \hat{\bmath{x}} +
 i \hat{\bmath{y}}
                       \right] e^{ -i Y }
                               \right\},
\label{eq:gf95_rhc}
\end{equation}
which dictates that, for GF95,
\begin{equation}
\hat{\bmath{e}}_R = ( \hat{\bmath{x}} + i \hat{\bmath{y}} )/\sqrt{2},
\label{eq:gf95_remark1}
\end{equation}
a form identical to that used by GKK73.

\subsubsection{IEEE compliance}
\label{remark2}

Although both GKK73 and GF95 are IEEE-compliant in their
description of circular polarization, and use identical
definitions of $\hat{\bmath{e}}_R$, the definitions of 
$\hat{\bmath{e}}_L$ are different; one can be obtained from the
other by multiplication by $-1$. From the point of view of
circular polarization, this difference is inconsequential. We note that the GKK73 definition,
$\hat{\bmath{e}}_L = (\hat{\bmath{x}}_L - i \hat{\bmath{y}}_L)\sqrt{2}$ may
be obtained from equation (\ref{eq:gf95_step3}) by setting $\delta=+\pi/2$,
for LCP radiation, to obtain
\begin{equation}
\bmath{E}(0,t) = {\cal E} \Re \left\{ 
  \hat{\bmath{x}}  \cos \omega_0 t -
  \hat{\bmath{y}}  \sin \omega_0 t
                              \right\}.
\label{eq:gf95_ieee_lhc}
\end{equation}
The wave in equation (\ref{eq:gf95_ieee_lhc}) is $y$-aligned at a new
start time given by $\omega_0 t_3 = -\pi/2$, and advancing it
through $\pi/2$ radians to $t_4=0$ yields an $x$-aligned field,
demonstrating clockwise rotation from the observer's viewpoint and
therefore IEEE-left-handedness. Insertion of the corresponding
phase factor, $e^{-i\pi/2}=-i$ into equation (\ref{eq:gf95_field}), with
equal real $x$- and $y$-amplitudes, then recovers the left-hand
unit vector used by GKK73. Note that this form of $\hat{\bmath{e}}_L$
is just the complex conjugate of $\hat{\bmath{e}}_R$, as defined
in equation (\ref{eq:gf95_remark1}).

Note that in the paragraph above, the starting time of the wave
was defined by $\omega_0 t_3 = -\pi/2 = \omega_0 t_1 -\pi$, where
$t_1$ is defined as in Section~\ref{testgf95} ($\omega_0 t_1 = \pi/2$).
The form of the LCP wave, at $z=0$, starting at $t_3$ is, from
equation (\ref{eq:gf95_field})
\begin{equation}
\bmath{E}(0,t-t_3) = {\cal E} \Re \left\{ 
                           \left[
  \hat{\bmath{x}}  -
 i\hat{\bmath{y}}
                           \right] e^{ -i \omega_0 (t-t_3) }
                     \right\},
\label{eq:lhc_t3}
\end{equation}
but to start it from the same time as the RCP wave tested in
Section~\ref{testgf95}, we eliminate $t_3$ in favour of $t_1$, noting
that this introduces a phase factor of $e^{-i\pi}=-1$, transforming
equation (\ref{eq:lhc_t3}) to
\begin{equation}
\bmath{E}(0,t-t_1) = {\cal E} \Re \left\{ 
                           \left[
 - \hat{\bmath{x}}  +
 i \hat{\bmath{y}}
                           \right] e^{ -i \omega_0 (t-t_1) }
                     \right\},
\label{eq:lhc_t1}
\end{equation}
which yields the form of $\hat{\bmath{e}}_L$ used in GF95, that is
$\hat{\bmath{e}}_L = (-\hat{\bmath{x}} + i \hat{\bmath{y}})/\sqrt{2}$.

\subsubsection{Definition of left-handed vector and scalar signature}
\label{remark3}

From Section~\ref{remark2}, we note that, for a given definition of a right-handed unit vector, two definitions of a left-handed vector are common: one is based on
the same initial orientation of the electric field vector (but different time origins), and the other is based on the same time (but different vector orientations for the LCPs and RCPs).
More importantly, the first type satisfies,
$\hat{\bmath{e}}_L = \hat{\bmath{e}}_R^*$, whilst the second satisfies
$\hat{\bmath{e}}_L = -\hat{\bmath{e}}_R^*$. This introduces a new
layer of complexity that does not affect the handedness of polarization,
but should be noted. 

We introduce here the scalar signature, $s$, which is the scalar product of a pair
of helical vectors in the spherical basis:
\begin{equation}
s = \hat{\bmath{e}}_R \cdot \hat{\bmath{e}}_L .
\label{eq:scalprod}
\end{equation}
$s$ is equal to either $+1$ for the first
type, where the vectors are complex conjugates, or $-1$ for the
second case, where they are anti-conjugate. In the tests considered
so far, GKK73 has $s=1$, but GF95 has $s=-1$.

\subsubsection{IAU compliance}

We have determined that polarization in GF95 follows the IEEE convention.
However, the values of $\delta$ used in \ref{testgf95}, combined
with the definition of Stokes $V$ in equation~(7) of that work imply that
GF95 used the non-IAU (LCP minus RCP) version of this quantity. A combination of 
non-IAU Stokes $V$ and the use of the $\sigma^\pm$ notation for transition
type results in a radiative transfer equation
for Stokes $V$ (equation~(24) of GF95) that is identical in form to that in GKK73
(who use IAU Stokes $V$, but whose $\pm$ subscripts are reversed from
our $\sigma^\pm$). The use of non-IAU Stokes $V$ by GF95 almost certainly
led to the incorrect statement regarding polarization handedness and field
orientation in Section~2.2 of that work. Note that our 
equation (\ref{eq:sigmaplus}) and equation (\ref{eq:sigmaminus}) may be obtained from
equation~(24) of GF95 by making the substitution
$V_{{\rm GF95}} = -V_{{\rm IAU}}$.

\section{The helicity of transitions}\label{helicitysection}
For GKK73 to comply with the IAU definition of Stokes~$V$, the right- and left-handed vectors used to represent the polarization of the radiation must have the following association with the positive and negative helicity vectors used to represent the response of the molecular density matrix:
\begin{equation}
\hat{\bmath{e}}_+ \rightarrow \hat{\bmath{e}}_R \;\; ; \;\;
\hat{\bmath{e}}_- \rightarrow \hat{\bmath{e}}_L \\ {\mathrm{[GKK73]}}
\label{eq:helpol_GKK73}
\end{equation}

In Section~\ref{descriptionfull}, we have considered the use of helical vectors to represent
the handedness of radiation polarization, noting that they are often written in the
form of positive and negative helicity as above. In order to derive equations
for the transfer of the Stokes parameters, we must also describe the
molecular response in terms of these same vectors. The molecular response
is used in two ways in the derivation of the transfer equations: once through
the macroscopic polarization of the medium that provides source terms for
the transfer equations, and again through the Hamiltonian that appears in the
evolution equations of the molecular density matrix. It turns out that the
Hamiltonian is adequate to fix the helicity of the various magnetic transitions,
and we will not consider the macroscopic polarization further.

By definition, the interaction Hamiltonian, comprising its off-diagonal elements
and resulting from the effect of the radiation field on the molecular electric dipole, has
elements of the form,
\begin{equation}
\hbar W_{a,b} = - \bmath{E} \cdot \hat{\bmath{d}}_{a,b},
\label{eq:wbar}
\end{equation}
where $a$ and $b$ are magnetic energy sublevels and $\hat{\bmath{d}}$ is the
dipole operator. For the sake of example, we choose a $\sigma^+$ transition, noting
that this has the lowest frequency in a Zeeman group (see Figure~1). Note that
the Zeeman group in GKK73 is the simplest non-trivial case with $m_{\rmn{upper}}$ having
the possible values $-1,0,1$ and an unsplit ground state with $m_{\rmn{lower}}=0$. Their
only allowed $\sigma^+$ transition is therefore the one from $m_{\rmn{upper}}=-1$ to the
ground state, and this is consistent with their equation (15), where the upper level
with $m=-1$ has the smallest diagonal Hamiltonian element of the three, and therefore
the lowest energy. Our specific form of equation (\ref{eq:wbar}) for the $\sigma^+$ transition
is therefore,
\begin{equation}
\hbar W_{-1,0} = - \bmath{E} \cdot \hat{\bmath{d}}_{-1,0} .
\label{eq:wbarspec}
\end{equation}
By comparison with equation (17) of GKK73 and our equation (\ref{eq:GKK73rl}), we see that the
dipole for the $\sigma^+$ transition is right-handed, and the dipole operator may
be written,
\begin{equation}
\hat{\bmath{d}}_{-1,0} = \hat{d} \hat{\bmath{e}}_R ,
\label{eq:dipole}
\end{equation}
which in the convention adopted by GKK73 corresponds to positive helicity
(see equation (\ref{eq:helpol_GKK73}) of this paper). This vector was originally written in a coordinate system
where the magnetic field lies along the $z$-axis, and a rotation matrix (equation
(19) of GKK73) was provided to rotate it onto the system where the $z$-axis
coincides with the direction of radiation propagation. In the present work it
is perfectly acceptable to align the systems, as in Figure \ref{molec}. Note that
equation (\ref{eq:dipole}) is consistent with this figure: molecules with an IEEE right-hand dipole interact
with only IEEE LCP radiation in the $\sigma^+$ transition, since
$s=1$, but a right-hand vector dotted onto itself is zero.

\subsection{Extension to Stokes $V$}

The formula derived for the molecular dipole in equation (\ref{eq:dipole}) is
only true when the magnetic field and radiation propagation directions
are aligned. If we reversed the magnetic field, whilst keeping the
propagation direction the same, our observer would then see
the molecular dipole for the same $\sigma^+$ transition
represented by an IEEE left-handed helical vector that
would interact with IEEE RCP radiation. By means of a rotation
matrix, the general case is represented in, say, the radiation propagation
frame as a linear combination of both left-handed and right-handed helical vectors with
coefficients that contain functions of the angle $\theta$ between the
propagation and magnetic field axes. For the case of the transfer of
Stokes $V$, the coefficients are functions only of $\cos \theta$.

Whatever detail is involved in obtaining it, the final radiative
transfer equation for IAU-compliant Stokes $V$ must agree with the
result derived in Section \ref{helicitysection} above: for a $\sigma^+$ transition: IEEE
LCP radiation is amplified when the magnetic field and propagation
axes are aligned ($\cos \theta = 1$). At this point, we note that GKK73
do not use the same $\sigma^+$ and $\sigma^-$ notation for magnetic
transitions as this paper: their definition follows their equation (46b)
on page~121, and states, `subscripts on the Stokes parameters 
distinguish among the three radiation bands by indicating the magnetic
sublevel of the upper state to which each couples'. With reference to Figure\,\ref{splittingfig}
we see that if $m_{\rmn{lower}}=0$ is the only available
lower state, as in GKK73, the upper state of the $\sigma^+$ transition
has $m_{\rmn{upper}}=-1$, corresponding to a subscript $-1$ for GKK73 --- i.e. their notation is reversed with respect to ours.

With the above note in mind, we select the simplest transfer equation for Stokes $V$ in GKK73 that provides the necessary information: their equation (52), line~2. Use of this equation requires that the magnetic field be sufficiently strong not only to provide a good quantization axis, but also to split the $\sigma^-$, $\pi$ and $\sigma^+$ transitions by considerably more than a Doppler (thermal and turbulent) width. The equation is also free of the complexities of saturation, which are not required to consider the sense of circular polarization. To place the equation in our $\sigma^-$ and $\sigma^+$ convention, we must change each $\pm$ symbol in the GKK73 equation with $\mp$ when it refers to a radiation quantity (Stokes $I$ and $V$ here), but not the one preceding the term in $2\cos \theta$, which is dictated by the field and propagation geometry discussed earlier. The results are our equation (\ref{eq:sigmaplus}) and equation (\ref{eq:sigmaminus}) for $\sigma^+$ and $\sigma^-$ respectively. Note that the above operation is not the same as a simple swap of helicities, which would swap all the $\pm$ to $\mp$ in the GKK73 equation and leave it invariant.

The discussion following equation (\ref{eq:sigmaplus}) and
equation (\ref{eq:sigmaminus}) in Section \ref{sigcorrespond} then gives us a predominance of negative
Stokes $V$ (IEEE LCP radiation) for a $\sigma^+$ (GKK73 negative) transition
when the magnetic and propagation axes are aligned ($\theta = 0$). This is
in accord with the discussion of the radiation-dipole interaction in
Section \ref{helicitysection}, with Figure \ref{molec}, and the classical Lorentzian derivation.

\section{Summary} 
\begin{figure}
\begin{center}
\renewcommand{\baselinestretch}{1.1}
\includegraphics[width=8cm]{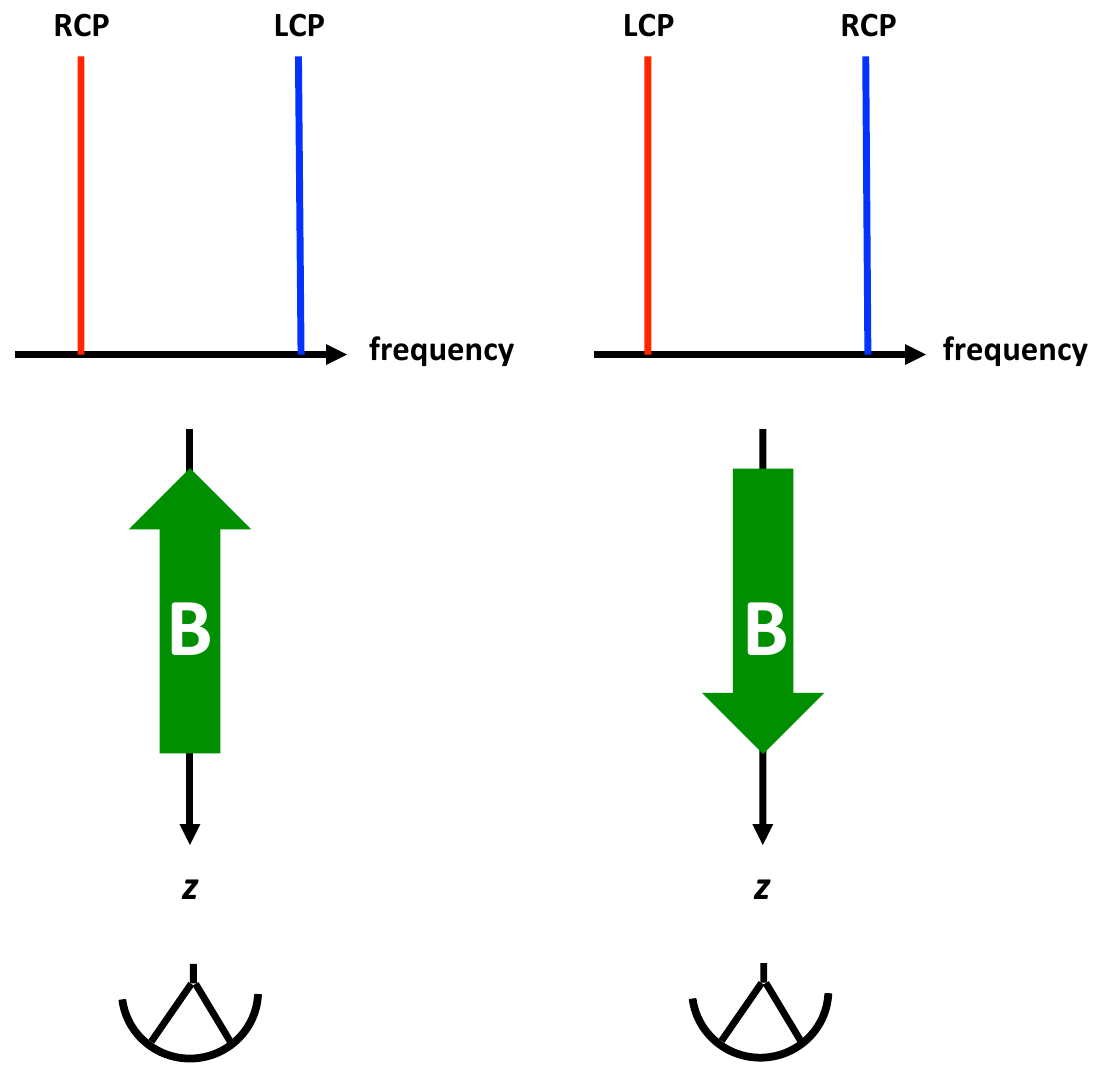}  
\caption{\small Summary of observed Zeeman profiles (top) and corresponding magnetic field directions relative to the observer (bottom). The IAU definition of Stokes $V$ and the IEEE definition of polarization handedness are assumed. Frequency increases from left to right.}
\label{summaryfig}
\end{center}
\end{figure}

We revisit the quantum mechanics and radiative transfer of maser emission under the conditions of Zeeman splitting and establish the correct field orientation for an observed spectrum. Adopting the IEEE convention for right-handed and left-handed circular polarization, and the IAU convention for Stokes $V$ (right-handed circular polarization minus left-handed circular polarization) we find (Figure \ref{summaryfig}):

\begin{itemize}
\item A magnetic field directed away from the observer will have right-hand circular polarization at a lower frequency (higher velocity) and left-hand circular polarization at a higher frequency (lower velocity). 
\item A magnetic field directed towards the observer will have right-hand circular polarization at a higher frequency (lower velocity) and left-hand circular polarization at a lower frequency (higher velocity). 
\end{itemize}

\noindent
 The results of our current analysis  are consistent with the classical Lorentzian derivations and mean that Zeeman splitting in the Carina-Sagittarius spiral arm, as measured from previous studies, should be interpreted as a field direction aligned away from the observer, and thus demonstrating a real field reversal in the interstellar medium.

\section*{Acknowledgments}
The authors thank the anonymous referee and the many people who have provided thoughtful discussions. MDG acknowledges the support of the STFC under grant ST/J001562/1.\\

\bibliographystyle{mn2e} \bibliography{UberRef}

\begin{thebibliography}{}

\bibitem[\protect\citeauthoryear{{Arfken}}{{Arfken}}{1970}]{arfken70}
{Arfken} G.,  1970, {Mathematical Methods for Physicists}.
Academic Press, 24/28 Oval Road, London, UK

\bibitem[\protect\citeauthoryear{{Babcock}}{{Babcock}}{1953}]{babcock53b}
{Babcock} H.~W.,  1953, ApJ, 118, 387

\bibitem[\protect\citeauthoryear{{Babcock} \& {Cowling}}{{Babcock} \&
  {Cowling}}{1953}]{babcock53a}
{Babcock} H.~W.,  {Cowling} T.~G.,  1953, MNRAS, 113, 357

\bibitem[\protect\citeauthoryear{{Boas}}{{Boas}}{1966}]{boas66}
{Boas} M.,  1966, {Mathematical Methods in the Physical Sciences}.
Wiley International, John Wiley \& Sons Inc., New York, USA

\bibitem[\protect\citeauthoryear{{Brown}, {Haverkorn}, {Gaensler}, {Taylor},
  {Bizunok}, {McClure-Griffiths}, {Dickey} \& {Green}}{{Brown}
  et~al.}{2007}]{brown07}
{Brown} J.~C.,  {Haverkorn} M.,  {Gaensler} B.~M.,  {Taylor} A.~R.,  {Bizunok}
  N.~S.,  {McClure-Griffiths} N.~M.,  {Dickey} J.~M.,    {Green} A.~J.,  2007,
  ApJ, 663, 258

\bibitem[\protect\citeauthoryear{{Davies}}{{Davies}}{1974}]{davies74}
{Davies} R.~D.,  1974, in {F.~J.~Kerr \& S.~C.~Simonson} ed., Galactic Radio
  Astronomy Vol.~60 of IAU Symposium, {Magnetic Fields in OH Maser Clouds}.
p.~275

\bibitem[\protect\citeauthoryear{{Davies}, {Verschuur} \& {Wild}}{{Davies}
  et~al.}{1962}]{davies62}
{Davies} R.~D.,  {Verschuur} G.~L.,    {Wild} P.~A.~T.,  1962, Nature, 196, 563

\bibitem[\protect\citeauthoryear{{Deguchi} \& {Watson}}{{Deguchi} \&
  {Watson}}{1990}]{deguchi90}
{Deguchi} S.,  {Watson} W.~D.,  1990, ApJ, 354, 649

\bibitem[\protect\citeauthoryear{{Dinh-v-Trung}}{{Dinh-v-Trung}}{2009}]{dinh09}
{Dinh-v-Trung} 2009, MNRAS, 399, 1495

\bibitem[\protect\citeauthoryear{{Edmonds}}{{Edmonds}}{1996}]{edmonds96}
{Edmonds} A.,  1996, {Angular Momentum in Quantum Mechanics}.
Princeton Landmarks in Mathematics and Physics, Princeton University Press

\bibitem[\protect\citeauthoryear{{Eisberg} \& {Resnick}}{{Eisberg} \&
  {Resnick}}{1974}]{eisberg74}
{Eisberg} R.,  {Resnick} R.,  1974, {Quantum Physics of Atoms, Molecules,
  Solids, Nuclei and Particles}.
Wiley International, John Wiley \& Sons Inc., New York, USA

\bibitem[\protect\citeauthoryear{{Elitzur}}{{Elitzur}}{1991}]{elitzur91c}
{Elitzur} M.,  1991, ApJ, 370, 407

\bibitem[\protect\citeauthoryear{{Elitzur}}{{Elitzur}}{1992}]{elitzur92}
{Elitzur} M.,  ed. 1992, {Astronomical masers} Vol.~170 of Astrophysics and
  Space Science Library

\bibitem[\protect\citeauthoryear{{Elitzur}}{{Elitzur}}{1993}]{elitzur93}
{Elitzur} M.,  1993, ApJ, 416, 256

\bibitem[\protect\citeauthoryear{{Field} \& {Gray}}{{Field} \&
  {Gray}}{1988}]{field88}
{Field} D.,  {Gray} M.~D.,  1988, MNRAS, 234, 353

\bibitem[\protect\citeauthoryear{{Field} \& {Gray}}{{Field} \&
  {Gray}}{1994}]{field94}
{Field} D.,  {Gray} M.~D.,  1994, A\&A, 292, 271

\bibitem[\protect\citeauthoryear{{Field} \& {Richardson}}{{Field} \&
  {Richardson}}{1984}]{field84}
{Field} D.,  {Richardson} I.~M.,  1984, MNRAS, 211, 799

\bibitem[\protect\citeauthoryear{{Fish}, {Reid}, {Argon} \& {Menten}}{{Fish}
  et~al.}{2003}]{fish03b}
{Fish} V.~L.,  {Reid} M.~J.,  {Argon} A.~L.,    {Menten} K.~M.,  2003, ApJ,
  596, 328

\bibitem[\protect\citeauthoryear{{Fish}, {Reid}, {Argon} \& {Zheng}}{{Fish}
  et~al.}{2005}]{fish05}
{Fish} V.~L.,  {Reid} M.~J.,  {Argon} A.~L.,    {Zheng} X.-W.,  2005, ApJs,
  160, 220

\bibitem[\protect\citeauthoryear{Fujia~Yang}{Fujia~Yang}{2010}]{yang10}
Fujia~Yang J. H.~H.,  2010, Modern Atomic and Nuclear Physics: Problems and
  Solutions Manual.
World Scientific Publishing Company

\bibitem[\protect\citeauthoryear{{Garcia-Barreto}, {Burke}, {Reid}, {Moran},
  {Haschick} \& {Schilizzi}}{{Garcia-Barreto} et~al.}{1988}]{garcia88}
{Garcia-Barreto} J.~A.,  {Burke} B.~F.,  {Reid} M.~J.,  {Moran} J.~M.,
  {Haschick} A.~D.,    {Schilizzi} R.~T.,  1988, ApJ, 326, 954

\bibitem[\protect\citeauthoryear{{Goldreich}, {Keeley} \& {Kwan}}{{Goldreich}
  et~al.}{1973}]{goldreich73part2}
{Goldreich} P.,  {Keeley} D.~A.,    {Kwan} J.~Y.,  1973, ApJ, 179, 111

\bibitem[\protect\citeauthoryear{{Gray}}{{Gray}}{2012}]{gray12}
{Gray} M.,  2012, {Maser Sources in Astrophysics}.
Cambridge University Press, Cambridge, UK

\bibitem[\protect\citeauthoryear{{Gray}}{{Gray}}{2003}]{gray03b}
{Gray} M.~D.,  2003, MNRAS, 343, L33

\bibitem[\protect\citeauthoryear{{Gray} \& {Field}}{{Gray} \&
  {Field}}{1994}]{gray94}
{Gray} M.~D.,  {Field} D.,  1994, A\&A, 292, 693

\bibitem[\protect\citeauthoryear{{Gray} \& {Field}}{{Gray} \&
  {Field}}{1995}]{gray95}
{Gray} M.~D.,  {Field} D.,  1995, A\&A, 298, 243

\bibitem[\protect\citeauthoryear{{Gray}, {Hutawarakorn} \& {Cohen}}{{Gray}
  et~al.}{2003}]{gray03a}
{Gray} M.~D.,  {Hutawarakorn} B.,    {Cohen} R.~J.,  2003, MNRAS, 343, 1067

\bibitem[\protect\citeauthoryear{{Green}, {McClure-Griffiths}, {Caswell},
  {Robishaw} \& {Harvey-Smith}}{{Green} et~al.}{2012}]{green12magmo0}
{Green} J.~A.,  {McClure-Griffiths} N.~M.,  {Caswell} J.~L.,  {Robishaw} T.,
  {Harvey-Smith} L.,  2012, MNRAS, 425, 2530

\bibitem[\protect\citeauthoryear{{Haken}, {Wolf} \& {Brewer}}{{Haken}
  et~al.}{2005}]{haken05}
{Haken} H.,  {Wolf} H.~C.,    {Brewer} W.~D.,  2005, {The Physics of Atoms and
  Quanta}

\bibitem[\protect\citeauthoryear{{Hamaker} \& {Bregman}}{{Hamaker} \&
  {Bregman}}{1996}]{hamaker96}
{Hamaker} J.~P.,  {Bregman} J.~D.,  1996, A\&AS, 117, 161

\bibitem[\protect\citeauthoryear{{Han} \& {Zhang}}{{Han} \&
  {Zhang}}{2007}]{han07}
{Han} J.~L.,  {Zhang} J.~S.,  2007, A\&A, 464, 609

\bibitem[\protect\citeauthoryear{{Jenkins} \& {White}}{{Jenkins} \&
  {White}}{1957}]{jenkins57}
{Jenkins} F.,  {White} H.,  1957, {Fundamentals of Optics, 3rd edition}.
McGraw-Hill, New York

\bibitem[\protect\citeauthoryear{{Jenkins} \& {White}}{{Jenkins} \&
  {White}}{1976}]{jenkins76}
{Jenkins} F.~A.,  {White} H.~E.,  1976, {Fundamentals of Optics 4th edition}.
McGraw-Hill, New York

\bibitem[\protect\citeauthoryear{{Kalkofen}}{{Kalkofen}}{1988}]{kalkofen88}
{Kalkofen} W.,  1988, {Numerical radiative transfer}.
Cambridge University Press, Cambridge, UK

\bibitem[\protect\citeauthoryear{{Landau}, P. \& B.}{{Landau}
  et~al.}{1982}]{landau82}
{Landau} E.~M.,  P. P.~L.,    B. B.~V.,  1982, {Quantum Electrodynamics}.
Butterworth-Heinemann, Oxford, UK

\bibitem[\protect\citeauthoryear{{Landi Degl'Innocenti} \& {Landolfi}}{{Landi
  Degl'Innocenti} \& {Landolfi}}{2004}]{landi04}
{Landi Degl'Innocenti} E.,  {Landolfi} M.,  eds, 2004, {Polarization in
  Spectral Lines} Vol.~307 of Astrophysics and Space Science Library

\bibitem[\protect\citeauthoryear{{Littlefield} \& {Thorley}}{{Littlefield} \&
  {Thorley}}{1979}]{littlefield79}
{Littlefield} T.~A.,  {Thorley} N.,  1979, {Atomic and nuclear physics. an
  introduction}.
Van Nostrand/Reinhold Co., New York

\bibitem[\protect\citeauthoryear{{Lothian}}{{Lothian}}{1957}]{lothian80}
{Lothian} G.,  1957, {Optics and its Uses. The Modern University Physics
  Series}.
Van Nostrand Reinhold

\bibitem[\protect\citeauthoryear{{Menegozzi} \& {Lamb} Jr.}{{Menegozzi} \&
  {Lamb}}{1978}]{menegozzi78}
{Menegozzi} L.~N.,  {Lamb} Jr. W.~E.,  1978, PhysRevA, 17, 701

\bibitem[\protect\citeauthoryear{{Reid} \& {Silverstein}}{{Reid} \&
  {Silverstein}}{1990}]{reid90}
{Reid} M.~J.,  {Silverstein} E.~M.,  1990, ApJ, 361, 483

\bibitem[\protect\citeauthoryear{{Robishaw}}{{Robishaw}}{2008}]{robishaw08}
{Robishaw} T.,  2008, PhD thesis, University of California, Berkeley

\bibitem[\protect\citeauthoryear{{Sommerfeld}}{{Sommerfeld}}{1954}]{sommerfeld54}
{Sommerfeld} A.,  1954, {Optics Lectures on Theortical Physics, Vol. IV}

\bibitem[\protect\citeauthoryear{{Stone}}{{Stone}}{1963}]{stone63}
{Stone} J.~M.,  1963, {Radiation and Optics}

\bibitem[\protect\citeauthoryear{{Surcis}, {Vlemmings}, {Curiel}, {Hutawarakorn
  Kramer}, {Torrelles} \& {Sarma}}{{Surcis} et~al.}{2011}]{surcis11}
{Surcis} G.,  {Vlemmings} W.~H.~T.,  {Curiel} S.,  {Hutawarakorn Kramer} B.,
  {Torrelles} J.~M.,    {Sarma} A.~P.,  2011, A\&A, 527, A48

\bibitem[\protect\citeauthoryear{{Van Eck}, {Brown}, {Stil}, {Rae}, {Mao},
  {Gaensler}, {Shukurov}, {Taylor}, {Haverkorn}, {Kronberg} \&
  {McClure-Griffiths}}{{Van Eck} et~al.}{2011}]{vaneck11}
{Van Eck} C.~L.,  {Brown} J.~C.,  {Stil} J.~M.,  {Rae} K.,  {Mao} S.~A.,
  {Gaensler} B.~M.,  {Shukurov} A.,  {Taylor} A.~R.,  {Haverkorn} M.,
  {Kronberg} P.~P.,    {McClure-Griffiths} N.~M.,  2011, ApJ, 728, 97

\bibitem[\protect\citeauthoryear{{Verschuur}}{{Verschuur}}{1969}]{verschuur69}
{Verschuur} G.~L.,  1969, ApJ, 156, 861

\bibitem[\protect\citeauthoryear{{White}}{{White}}{1934}]{white34}
{White} H.~E.,  1934, {Introduction to atomic spectra}

\bibitem[\protect\citeauthoryear{{Young} \& {Freedman}}{{Young} \&
  {Freedman}}{2004}]{young04}
{Young} H.,  {Freedman} R.,  2004, {University Physics - 11th Edition}.
Pearson-Addison-Wesley

\bibitem[\protect\citeauthoryear{{Zeeman}}{{Zeeman}}{1897}]{zeeman1897}
{Zeeman} P.,  1897, ApJ, 5, 332

\bibitem[\protect\citeauthoryear{{Zeeman}}{{Zeeman}}{1913}]{zeeman1913}
{Zeeman} P.,  1913, Koninklijke Nederlandse Akademie van Wetenschappen
  Proceedings Series B Physical Sciences, 16, 155

\end{thebibliography}

\appendix

\section{A simple test}

The scalar or dot product of pairs of helical vectors in the spherical basis has been introduced in Section~\ref{remark3} to define the scalar signature of
a basis. The dot product may also be used as a very simple test to determine conformity of any basis to the IEEE convention now that we
have a vector, $\hat{\bmath{e}}_R = ( \hat{\bmath{x}} + i \hat{\bmath{y}} )/\sqrt{2}$, that we know is IEEE-compliant.

Unlike the real unit vectors of all common axis systems (e.g. cartesian), a helical basis vector dotted onto itself yields the result zero. Compliance with the IEEE convention can therefore be tested by taking the right-hand vector for test, $\hat{\bmath{e}}_{R?}$, and calculating the dot product, $\hat{\bmath{e}}_R \cdot \hat{\bmath{e}}_{R?}$. If the result is zero, the tested system is IEEE-compliant; if it is $\pm 1$, the tested system is optics-compliant. All the works below were tested in this way. Table \ref{test_results} presents a summary table of the test results. Note that it has been assumed throughout that the listed authors intended to follow the IAU convention unless clearly stated otherwise, and in some cases conformity to the IEEE convention on the handedness of polarization depends on this assumption by the present authors. 

\begin{table}
 \caption{Electric field polarization conventions apparently used
in works discussing maser polarization and in selected general texts.}
 \label{test_results}
 \begin{tabular}{@{}lcc}
  \hline
  Author(s)$^{1}$ & Convention$^{2}$ & Signature$^{3}$ \\
  \hline
   GKK73 & IEEE & $+1$ \\
  GF95$^{4}$ & IEEE & $-1$ \\
  DW90 & IEEE & $-1$ \\
    E91 & optics         &$+1$ \\
 DvT09 & IEEE & $-1$ \\
   G12 & optics & $-1$ \\
  \hline
 \end{tabular}

 \medskip
$^{1}$ See main text for definitions of shorthand notation.\\
$^{2}$ IEEE corresponds to $\hat{\bmath{e}}_R \cdot \hat{\bmath{e}}_{R?}=0$;  optics corresponds to $\hat{\bmath{e}}_R \cdot \hat{\bmath{e}}_{R?}= \pm1$.\\
$^{3}$ From evaluation of $\hat{\bmath{e}}_R \cdot \hat{\bmath{e}}_L$, as discussed in Section \ref{remark3}.\\
$^{4}$ The same convention was used in the following works:
\citet{gray94}, \citet{field94}, \citet{gray03b}, \citet{gray03a}. Stokes $V$ is defined contrary to the IAU convention.\\
\end{table}

\subsection{\citet{deguchi90}}

This work (DW90 for short) is concerned mostly
with linear polarization, but certainly contains sufficient information
to straightforwardly determine the convention used for circular
polarization. We note that the paper clearly states that the choice
of basis vectors is different from that used by GKK73 in order to be
consistent with more general work concerned with transitions other
than $J=1-0$; specific reference is made to
\citet{edmonds96}.

In order to be IAU compliant, equation~(A20) of DW90 requires
that the electric field component labelled ${\cal E}_-$ be RCP.
However, we see from the electric field definition, equation~(A1), that
${\cal E}_-$ is actually the coefficient of 
$-\hat{\bmath{e}}_+$, that is there is a sign swap between the field
and its associated helical vector. From the above argument, and their
equation~(A4a), DW90 therefore require the
spherical basis,
$\hat{\bmath{e}}_R = ( \hat{\bmath{x}} + i \hat{\bmath{y}} )/\sqrt{2}$,
$\hat{\bmath{e}}_L = ( -\hat{\bmath{x}} + i \hat{\bmath{y}} )/\sqrt{2}$.
This is IEEE compliant, with scalar signature equal to $-1$, and
therefore in the same convention as GF95.

\subsection{\citet{elitzur91c}}

This work (E91 for short) includes \citet{elitzur91c} and subsequent works \citep{elitzur92,elitzur93}. Its purpose was to extend the work of GKK73 from $J=1-0$ to arbitrary rotational transitions. In this work $E^{+}$ corresponds to LCP, and it is assumed $E^{+}$ is the coefficient of $\hat{\bmath{e}}_+$ and equivalent to $\hat{\bmath{e}}_L$. Under these assumptions the vectors are optics compliant and the scalar signature is $+1$. However, it should be noted that Stokes $V$ is defined contrary to IAU, with Stokes $V$ = LCP -- RCP.

\subsection{\citet{dinh09}}

The work considered here (DvT09 for short) applied up-to-date
computing power to the polarization problem, with conclusions that
supported the standard model of polarized maser propagation. The
definition of Stokes $V$ is the same as in DW90 (DvT09 equation~(7)),
so the electric field component marked with negative helicity must
be RCP. The electric field definition (DvT09 equation~(4))
also agrees with DW90 so we must equate
$\hat{\bmath{e}}_R = -\hat{\bmath{e}}_+$ and
$\hat{\bmath{e}}_L = -\hat{\bmath{e}}_-$. 
By inspection of DvT09 equation (1), we can see that the definition
of $\hat{\bmath{e}}_R$ agrees with DW90 and GF95. However, the
definition of $\hat{\bmath{e}}_L$ appears to be simply a negated
version of $\hat{\bmath{e}}_R$. Presumably this is just a
typographical error, and the symbol before $i$ should be $\pm$
rather then $+$. Assuming this is so, DvT09 is 
IEEE-compliant.

\subsection{\citet{gray12}}

In the above work (G12 for short) the Stokes parameters are
set out on page~277. They have Stokes $V$ as positive helicity
field components subtracted from negative helicity components
as in DW90 and DvT09. Therefore, to conform with the IAU convention,
negative helicity components must be RCP. However, as in GKK73,
but unlike DW90 and DvT09, the negative helicity 
(right-handed) field-component is
the amplitude associated with the negative helicity unit vector,
as required by equation~7.153 of G12 on page~270.
We therefore make the association, $\hat{\bmath{e}}_R = \hat{\bmath{e}}_-$
in this case. From the definitions printed just above equation~7.153,
and the discussion above, G12 defines the right- and left-handed
unit vectors as,
$\hat{\bmath{e}}_R = (\hat{\bmath{x}} - i \hat{\bmath{y}})/\sqrt{2}$ 
and
$\hat{\bmath{e}}_L = -(\hat{\bmath{x}} + i \hat{\bmath{y}})/\sqrt{2}$.
This combination is optics-compliant with a scalar signature of $-1$.
\citet{gray12} is therefore internally self-consistent in having a field
pointing away from the observer in W3(OH).

\label{lastpage}

\end{document}